# Empirical Validation of Network Learning with Taxi GPS Data from Wuhan, China

Susan Jia Xu, Qian Xie, Joseph Y. J. Chow, and Xintao Liu

*Abstract*—In prior research, a statistically cheap method was developed to monitor transportation network performance by using only a few groups of agents without having to forecast the population flows. The current study validates this "multi-agent inverse optimization" method using taxi GPS trajectories data from the city of Wuhan, China. Using a controlled 2062-link network environment and different GPS data processing algorithms, an online monitoring environment is simulated using the real data over a 4-hour period. Results show that using only samples from one OD pair, the multi-agent inverse optimization method can learn network parameters such that forecasted travel times have a 0.23 correlation with the observed travel times. By increasing to monitoring from just two OD pairs, the correlation improves further to 0.56.

## I. Introduction

**M**ANY studies have illustrated the importance to accurately and precisely measure the attributes of an urban transport system. Due to the rise of Big Data and Internet of Things, there are numerous machine learning methods to measure attributes of the transport system. Chow [1] provides an overview of these techniques including several applications like Allahviranloo and Recker [2] for activity pattern prediction; Cai et al. [3] for short-term traffic forecasting; Luque-Baena et al. [4] for vehicle detection; Lv et al. [5] for traffic flow prediction; and Ma et al. [6] for network congestion prediction. However, generic machine learning techniques are not specifically designed to exploit the unique structure of urban transport networks.

As a result, in recent years a theory of inverse problems (see [7]) have emerged to capture network structure, dubbed "inverse transportation problems" by Xu et al. [8]. If a conventional model $M$ that transforms a set of parameters $\theta$ to a set of outputs $X$ as $X = M(\theta)$, then the inverse model deals with estimating the parameters $\hat{\theta}$ based on observed outputs $x$ as $\hat{\theta} = M^{-1}(x)$. Many types of inverse transportation problems have been proposed in the literature: inverse shortest path [9]; inverse linear programs for an assortment of transportation problems [10]; link capacities in minimum cost flow problems [11]; inverse vehicle routing problems [12;13]; general inverse variational inequalities for equilibrium models [14]; and route choice [15].

Despite the growing literature, inverse transportation problems are designed to take a system level model and estimate parameters of that model from sample data. This is problematic because congested systems require estimation of population attributes like flow in order to quantify congestion effect parameters because the more congested the system the more of an outlier it becomes. Ma et al. [6] is an example of this type of effort, using deep Restricted Boltzmann Machines and Recurrent Neural Networks to estimate population-level flows based on taxi trajectory sample data. Another challenge is the lack of consideration of behavioral mechanisms. Many inference models, particularly those belonging to "network tomography" (see [16; 17]), explains the state of the system from the data but do not explain the behavioral mechanisms like route choice on the flow attributes. Inverse transportation problems like Güler and Hamacher [11] result in NP-hard problems that are not scalable to practical size networks.

Xu et al. [8] recently proposed a theory based on multi-agent inverse optimization (MAIO) in which the capacity effects of a network are inferred using sampling multi-agent inverse transportation problems instead of solving a single system-level problem. An agent in this context refers to either an individual, or collective individuals, that makes travel decisions to interact with a virtual environment (see [42,43]). This is possible under the assumption that the agents sampled exhibit behavioral characteristics like route choice preferences. The model infers congested links' capacity dual variables by using sample of agents' inverse shortest path problems. This method quantifies and explains the congestion in the network, i.e. not just how congested it is, but how much each link's congestion impacts the rest of the network (which is more interpretative than methods like [6], i.e. we are not just predicting travel times in

Submitted on November 11, 2019 for review. This study was conducted with support from the NSF CAREER grant, CMMI-1652735. One of the authors (Xintao Liu) was supported by The Hong Kong Polytechnic University Start-up Research Fund Program: [Grant Number 1-ZE6P]; Area of Excellence: [Grant Number 1-ZE24].

Susan Jia Xu was with the New York University, New York, NY 11201 USA (e-mail: jx731@nyu.edu). She is now with the San Diego Association of Government (SANDAG), San Diego, CA 92101 USA

Qian Xie was with Tsinghua University, Beijing, China. She is now with the Department of Civil & Urban Engineering, New York University, New York, NY 11201 USA (e-mail: qx463@nyu.edu).

Joseph Y.J. Chow is now with the Department of Civil & Urban Engineering, New York University, New York, NY 11201 USA (e-mail: joseph.chow@nyu.edu).

Xintao Liu is with the Hong Kong Polytechnic University, Hong Kong, China. (e-mail: xintao.liu@polyu.edu.hk).

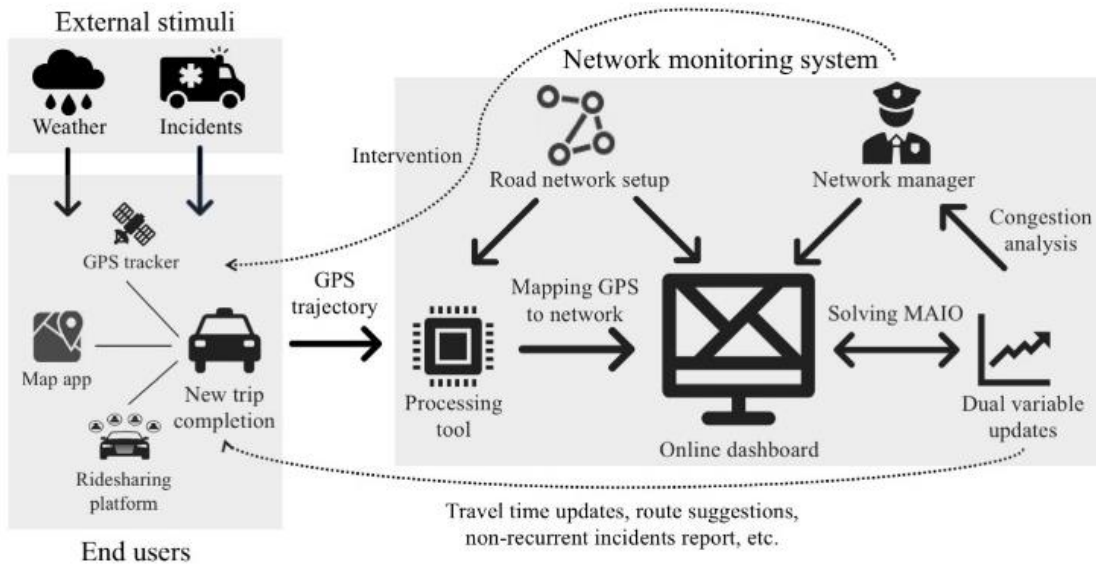

Fig. 1. Network monitoring architecture using method from [8].

the network but we are explaining the travel time in a part of the network due to congestion in another part). A test of the model using queried data from a highway network in Queens, NY, demonstrated the methodology in being able to monitor the network over time and use samples to update the network's link capacity effects.

That earlier work developed the theory for the method and validated it using a controlled data setting with queried routes across 56 different potential origin-destination (OD) pairs in Queens, NY. While the time travel times correspond to real data, (1) the sampling of OD pairs over time is not based on real demand for information and (2) the realized route choices are assumed to be the same as Google queries as opposed to realized choices collected from the field. In practice, this method would need to serve a system design illustrated in **Fig. 1**. The success of the system design depends on where the sample data from end users are coming from (e.g. crowd-sourced participants like Google Waze or GPS data from regulated taxis, or both?). If route data collected from the field only corresponds to certain OD demand, which OD demands sufficiently monitor the network, and how many samples are needed?

The contribution of this study is an empirical study to validate these implementation issues: how effectively can we use the MAIO method from [8] to monitor a 2062-link network in Wuhan, China, using field data of realized route choices collected from a controlled number of OD pairs? In [8], an experiment was conducted using externally queried routes made periodically in a lab to evaluate the performance of the algorithm. Because the data source for the monitoring of the network is from an existing sensor (Google), it only validates the sensitivity of the algorithm to state changes. *It does not validate the methodology's effectiveness at both sensing and inference.* This current study uses real trajectories from sampling rates that reflect actual travel conditions made within the network by travelers experiencing the congestion. It tests the effectiveness of using sampled data from two OD pairs in the network in acting as both sensors and inference mechanisms. By proving the effectiveness using only two OD pairs, the study gives credence to larger monitoring systems that can make use of multiple OD pair trajectories. As **Fig. 1** shows, a monitoring system requires either processing of GPS trajectory data to map it to a network data structure or a data collection device that automatically outputs location data in the network data structure. Since, many data sources are only provided from GPS data, we also propose mapping algorithms to match the location data to network data structures.

The remainder of the paper is organized as follows. Section 2 reviews studies using taxi trajectory data, and network attributes (i.e. travel time) estimation method, including the methodology from Xu et al. [8]. Section 3 presents the experiment design, data preparation, and on-line system simulation set-up steps. The data processing algorithms used in this experiment are introduced in this section. Section 4 discusses the experiment results. Section 5 concludes.

## II. Literature Review

In a real-world setting, one way to obtain traffic data in a network is from GPS-equipped vehicles. Jenelius and Koutsopoulos [18] called this kind of data as "floating-car" data, where vehicles with GPS equipment installed record their location and speed at fixed time intervals ranging from a few seconds to minutes. As an important component of the urban transport system, taxi offers an all-weather, convenient, comfortable, and personalized travel service for the urban residents, as well as plays a key role in the urban passenger mobility development [19]. Taxi GPS trajectories data has been widely used in transportation research, including travel time estimation [20; 21] and travel behavior analysis [22 – 25], or for inferring travel momentum in a city [26; 27]. Such probe data is useful for evaluating recurrent and non-recurrent incidents to mitigate their impacts on traffic. Examples include [39] – [41], where GPS data may be combined with other sensors and count data to monitor the effects of incidents over time.

Based on the taxi GPS traces and data mining technology, we can obtain the experienced taxi driver's route choice behavior in real time and provide guides of shortest path optimal choice for general public [28]. There are extensive studies of taxi operations and route choice analysis; however, in this case study, we do not aim at analyzing taxi driver's practical travel behavior or their activity analysis. The taxi trajectory data is used as sampled heterogeneous agent information to test the MAIO method with a large network.

There are numerous network congestion inference methods using different traffic data. Early efforts in network tomography from Vardi [16] and Tebaldi and West [17] proposed methods to estimate flow distributions from observed link count data. Other studies have sought to integrate route choice behavioral mechanisms in the estimation [29; 30; 31; 32; 33]. Deep learning models have been proposed for capturing network congestion [6]. System level inverse transportation problems have been proposed to estimate path flows of taxis to have consistent network congestion characteristics [34], who suggested adding a term in the optimization objective function penalizing the travel time between the observation and the sum of link travel times along the path.

The MAIO method from Xu et al. [8] introduces the dual variables in the objective function. The values of dual indicate the change of network state (such as congestion effects) in the form of travel time. While the method does not explicitly model traffic flow dynamics, it is implicitly capturing them through the estimation of the effects using a linear optimization model (see [44, 45]).

Xu et al. [8] defined a network $G(N, A)$ that receives observations from a sample $P$ of agents seeking to travel from an origin node $r_i \in N$ to a destination node $s_i \in N$, $\forall i \in P$. In the MAIO method, we assume each agent $i \in P$ is rational traversing a network modeled as a capacitated multicommodity problem shown in matrix form in Eq. (1) – (4), where $c_a, a \in A$, is the free flow link cost, $x_m$ is the flow of OD pair $m \in M$, $A$ is the node-link incidence matrix, $b_m$ is $+q_m$ at the source node for OD pair $m$, $-q_m$ at the sink node, and 0 otherwise. $u_a, a \in A$, is the link capacity. In the case where there are more factors to choosing a route (e.g. in multimodal networks) a route choice model can be estimated to determine a generalized cost function to replace $c_a$.

$$\min_x \sum_m c^T x_m \tag{1}$$

Subject to

$$A x_m = b_m, \qquad \forall m \in M \tag{2}$$

$$\sum_{m \in M} x_m \le u \tag{3}$$

$$x_m \ge 0, \qquad \forall m \in M \tag{4}$$

Solution of this problem can involve a decomposition into a restricted master problem to determine the dual variables $w_a$ corresponding to link capacities $u_a$. Based on the dual variables, subproblems for each OD pair can then be solved in unbundled form as unconstrained shortest path problems shown in Eq. (5) – (7), where $b$ is a vector of either +1 at the origin, -1, at the destination, and 0 otherwise. The notation $\phi$ represents the shortest path operator, where $\phi^{-1}$ is the inverse operator. The dualized link costs are cost $\bar{c}_a$, i.e. $\bar{c}_a = c_a + w_a$. When there's no congestion, $w_a = 0$. When there is sufficient congestion to cause behavioral change in route choice, $w_a > 0$.

$$\min_y \phi = (c + w)^T y \tag{5}$$

Subject to

$$A y = b \tag{6}$$

$$y_a \in \{0, 1\}, \qquad a \in A \tag{7}$$

The MAIO exploits this structure to estimate each agent's perception of $w_a$, denoted as $w_{a,i}$. In the inverse problem, we observe $y_i^*$ for each agent $i \in P$. If the path chosen is the shortest path according to free flow conditions, then $w_{a,i} \ge 0$ on the path chosen. If another path is chosen, the $w_{a,i}$ for the free flow shorter path needs to be increased. Increasing them optimally to suit each agent is an inverse shortest path problem shown in Eq. (8) – (12) as derived as a linear program from Ahuja and Orlin [10], where $v_i$ are the unbounded node potentials. This problem assumes prior dual variables for each link capacity constraint are available, $\bar{w}$. The objective is to minimally perturb from the priors to obtain a new dual variable $w_i = \bar{w} - e_i + f_i$ for agent $i \in P$ based on observing their chosen route $y_i$ (Eq. (8)), subject to weak duality (Eq. (9)), strong duality (Eq. (10)), capacity dual variable feasibility (Eq. (11)), and non-negativity constraints (Eq. (12)). $w_i^* = \phi_i^{-1}(g_i, \bar{w}, y_i^*)$ is a function of the OD locations of the agent represented as their graph parameters $g_i$, the prior, and the chosen path $y_i^*$.

$$\min_{e_i, f_i, v_i} \phi_i^{-1} = e_i + f_i \tag{8}$$

Subject to

$$A^T v_i \le c + \bar{w} - e_i + f_i \tag{9}$$

$$b^T v_i = (c + \bar{w} - e_i + f_i)^T y^* \tag{10}$$

$$e_i - f_i \le \bar{w} \tag{11}$$

$$e_i, f_i \ge 0 \tag{12}$$

In an online setting, we assume the population $P$ arrives sequentially over time. In that case, the value of $\bar{w}$ is obtained from a previous agent observation $i - 1$ as $\bar{w} = w_{i-1}^*$ and used to feed a current observation $i$ to update $w_i^*$. This is summarized in **Algorithm 1**. It makes use of inverse shortest path problems, which can be solved as computationally efficient as other shortest path algorithms (see [8]), which are known to be solvable with $O(n \log n)$ efficiency. The algorithm therefore uses trajectory data as sensors and the route decisions to support the inference.

---

**Algorithm 1: online learning algorithm to update system capacity dual variables**

---

0.  Given: a prior (obtained from a system) $w_0^* = 0$.
1.  For each new arrival $i$,
    a.  Set $\bar{w} = w_{i-1}^*$.
    b.  Solve an inverse shortest path problem with augmented link costs, $w_i^* = \phi^{-1}(g_i, \bar{w}, x_i^*)$.

---

In [8], the only validation done is on the sensitivity of the algorithm to state changes in the network, using externally queried (Google) routes made periodically in a lab. It does not validate the methodology's effectiveness at both sensing and inference, which requires real in-situ data and a carefully designed experiment that compares that data with inference sensed using the same data.

No tests have been conducted to validate the correlation between output dual variables acting as both inference and sensor from real data. Without this, there is no empirical proof that the method itself works in that manner, which is necessary for traffic network managers to consider adopting the technology.

## III. PROPOSED EXPERIMENT DESIGN

We test the multi-agent IO approach using real taxi data from Wuhan, China. Having observed travel times of the taxis along links, can we demonstrate the existence of correlations with the travel times generated from our monitored link dual variables and the realized travel times under a simulation of an online operation? That is the research question that needs to be addressed by this empirical study. We consider the following criteria to evaluate:

i.   Comparison of the predicted route and the actual route chosen;
ii.  Compute the correlation between real travel times and estimated travel time.

Based on these two criteria, we design an experiment involving multiple time interval observations and evaluate the performance of the MAIO method. Our goal is to show that even with limited OD data, we can see improvement in accuracy of the monitoring system as we go from one OD pair sampling to two OD pair sampling, because additional information will only improve the output.

### A. Data preparation: network

The data consists of taxi GPS trajectories from Wuhan, China. Wuhan is the capital of Hubei province in China, with a population of 11 million (9th largest city in China). In the mega-city level congestion list [36], Wuhan was ranked 10th. As shown in **Fig. 2**, Wuhan is composed of thirteen districts separated by the Yangtze River. Hanyang and Wuchang districts form part of the urban core of Wuhan, with both served by metro lines.

An urban transport network from Wuhan, China overlaid on OpenStreetMap is shown in **Fig. 3**. The network attributes are available on GitHub [38]. Sampled data of free flow link travel times ("FF time") are presented in **Table I**. There are 2,833 links and 855 nodes in the extracted urban network. The network is designed to monitor two origin-destination (OD) pairs (see red dots in **Fig. 3**): *Zhongjiacun Station* (Line 4 and Line 6, Hanyang District) to *Wuchang Rail Station* (Line 4 and Line 7, Wuchang District), and *Zhongjiacun Station* to *Pangxiejia Station* (Line 2 and Line 7, Wuchang District), which are selected from hot spots of pick-ups and drop-offs in the city.

TABLE I
SAMPLE OF LINK ATTRIBUTES FOR THE STUDY NETWORK

| Link ID | Start Node ID | End Node ID | Free Flow Time_sec |
| ( $a$ ) | ( $O$ ) | ( $D$ ) | ( $c$ ) |
|---|---|---|---|
| 9 | 12 | 1500 | 16.04 |
| 10 | 12 | 588 | 9.34 |
| 13 | 20 | 1516 | 7.03 |
| 14 | 20 | 1504 | 37.43 |
| 15 | 20 | 28 | 5.06 |
| 16 | 22 | 237 | 17.09 |
| 17 | 22 | 1298 | 14.14 |
| 18 | 22 | 17 | 2.66 |
| 19 | 28 | 20 | 5.20 |

The four-hour taxi trajectory data on May 6th, 2014, from 5AM to 9AM, is used for the test. This period is the peak of the day as shown in the time of day trip times in **Fig. 4**. The period analyzed represents non-stationary trip volumes so if the method works for this it is trivial to extend this to other time periods as well.

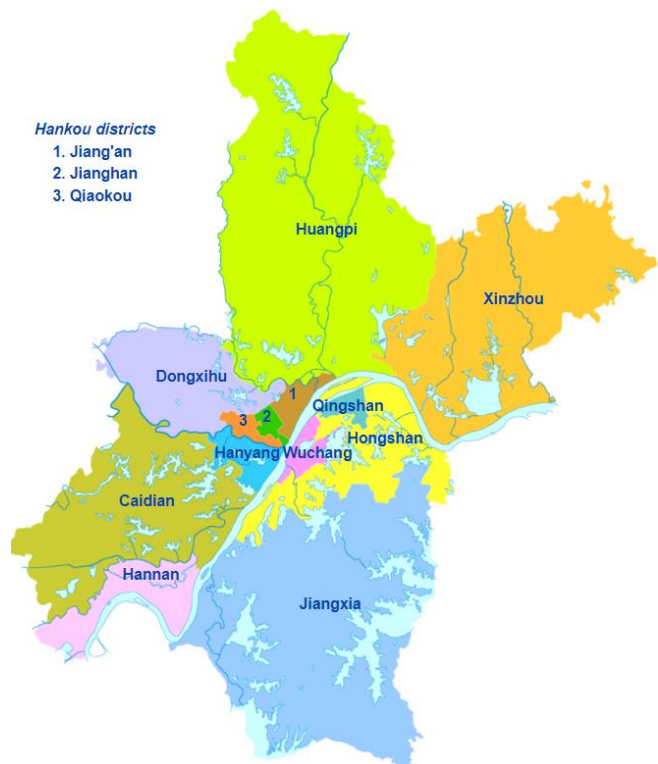

Fig. 2. Wuhan districts (source: [37]).

Algorithms for path reconstruction using GPS coordinates are summarized in **Algorithm 2 – Algorithm 5**. The processed path data, along with the network information and network learning code, are all located in the GitHub [38] site.

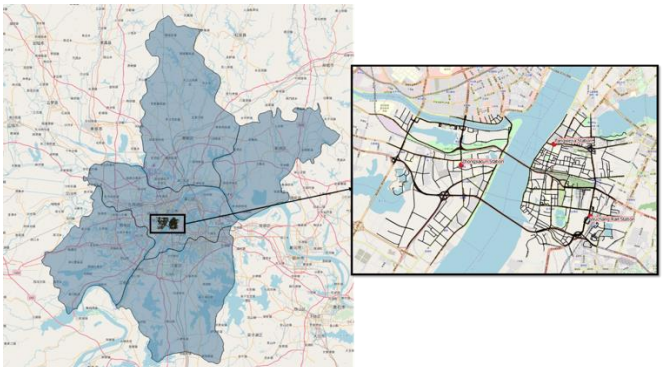

Fig. 3. Study urban transport network in Wuhan, China.

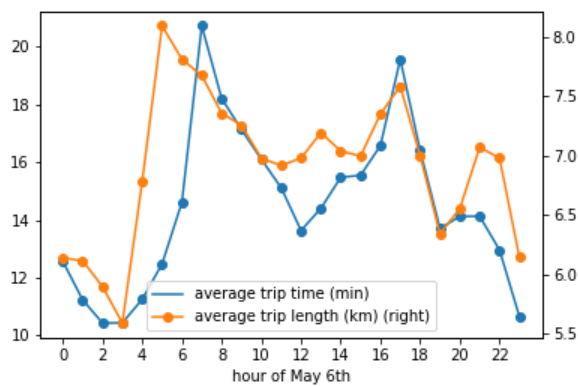

Fig. 4. Average trip time in minutes (left axis) and distance in km (right axis) by time of day.

### B. Data preparation: simulation setup

The network is initiated under free flow condition. The multi-agent IO approach keeps updating the travel cost for the whole network every time new path information is obtained (e.g. from the taxi GPS records). When a new path is obtained, we update the effect that the link capacities have on the path using Algorithm 1 such that the observed path is perceived by the agent to be optimal. We monitor and update the system over 4 hours in this case.

The following steps are taken for the experiment.
1.  Initiate with values of dual variables equal to zero for all links in the urban transport network in Wuhan, China.
2.  Starting at 5:00AM, and every 5 minutes thereafter until 9:00AM,
    a.  For all the trajectories that arrived in that period, identify origin and destination (OD) pairs.
    b.  Run the path reconstruction algorithms (see **Algorithm 2 – Algorithm 5**) to get real-time travelers' choices for each of the OD pairs (in this step, the traveler's choice is assumed as the shortest path).

c.  Compare the predicted route and the actual route chosen.
d.  Run **Algorithm 1** to update the link dual variables based on the reconstructed path.
e.  Compute the correlation between real travel times and estimated travel time.

As congestion occurs in the network, the effects of the capacity on shifting routes (see **Fig. 5** for an illustration of these changes over different time intervals) should be recognized by the network learning algorithm. The dual variables should reflect links that become more congested with binding capacity effects that result in route diversions. The magnitudes of the dual variables should give a relative measure of the insufficient capacity in the link with respect to other links.

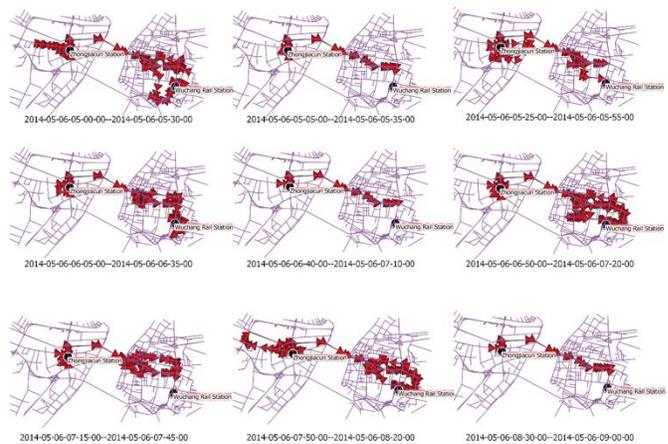

Fig. 5. Sample of route diversions for one OD (from Zhongjiacun Station to Wuchang Rail Station).

### C. Data preparation: taxi trajectories

There are about 8,200 taxis operating over 16 hours a day in the city. The dataset used in this test, referred to as the "City of Wuhan Taxi" (COWT) data, contains GPS trajectories of all registered taxis in Wuhan, China. Each GPS entry has information including taxi id, longitude/latitude, time stamp, instantaneous velocity and heading, the operation and occupancy status, as shown in **Table II**. The minimum interval between two data points is around 15 seconds, and the maximum one is 2 minutes.

TABLE II
SAMPLE DATA OF COWT

| ID[1] | Timestamp[2] | Longitude | Latitude | Angle[3] | Speed[4] | Operation | Status[5] |
|---|---|---|---|---|---|---|---|
| 10287 | 5/4/2014 23:59 | 114.300472 | 30.557818 | 64 | 20 | Operate | 0 |
| 12448 | 5/4/2014 23:59 | 114.137636 | 30.600324 | 55 | 15 | Operate | 0 |
| 4864 | 5/4/2014 23:59 | 114.214882 | 30.571331 | 94 | 51 | Operate | 1 |
| 8695 | 5/4/2014 23:59 | 114.320283 | 30.636952 | 0 | 0 | Operate | 0 |
| 8538 | 5/4/2014 23:59 | 114.298862 | 30.602568 | 0 | 0 | Operate | 1 |
| 2034 | 5/4/2014 23:59 | 114.197638 | 30.558353 | 0 | 0 | Operate | 0 |
| 6700 | 5/4/2014 23:59 | 114.323372 | 30.521492 | 0 | 1 | Operate | 0 |
| 5620 | 5/4/2014 23:59 | 114.415055 | 30.478973 | 184 | 54 | Operate | 0 |
| 10179 | 5/4/2014 23:59 | 114.282767 | 30.612157 | 190 | 25 | Operate | 0 |

1-ID: Taxi ID; 2-Timestamp: sampling time; 3-Angle: North (0) and South (180); 4-Speed: kilometer per hour; 5-Status: occupied (1) and vacant (0)

The occupancy status associated with each GPS record, which indicates whether there is a passenger in the taxi, is the input to trip segmentation (e.g. the process if dividing taxis trajectories into occupied and vacant trips). A simple rule-based filter can identify unrealistic short occupied trips and fix sudden flips in occupancy status. The processing of taxi trajectories includes the following tasks:

1. GPS points are mapped to the road network based on their coordinates using QGIS
2. Outliers in the trajectory are filtered based on a simple rule, i.e., the speed associated with each GPS points cannot be greater than 120 km/h. Due to the data loss, some taxis have longer time interval between two consecutive GPS points.
3. Trajectories are split into sequences using a time gap threshold of 300 seconds (5-min).

### D. Data preparation: hotspot identification

Trajectories are split into occupied and vacant trips primarily based on the observed occupancy status. The distribution of trip origins and destinations on the day of May 6th, 2014, is reviewed, and it is expected that pick-ups and drop-offs are more likely to occur in hot spot areas. Hence, heatmaps of taxi pick-ups and drop-offs are created in QGIS as shown in **Fig. 6(a)**. The heatmaps show where there is a high concentration of pick-ups and drop-offs, respectively. The hotspots are identified as clusters in **Fig. 6(b)** using the Hotspot Analysis Plugin in QGIS, which are extracted from the heatmaps. In this case, we select the metro station – *Zhongjiacun* as the origin, and another two metro stations: *Wuchang Rail Station* and *Pengxiejia* as two destinations to have a controlled setting for this experiment.

### E. Data preparation: trip extraction

We define a **trip** as a journey made by a taxi from picking up passengers to dropping off passengers. Since COWT data provides us with the "status" information, we regard a taxi is picking up passengers if the "status" changes from 0 to 1 and dropping off passengers if the "status" changes from 1 to 0. The phrase **trip trajectory** refers to the GPS records generated during part of a trip, which means the status is always 1 in a trip trajectory. Once the OD pair for each experiment is determined, we select those trip trajectories with a starting point close to (within a specific threshold $\alpha$, say 500 meters) the origin and an ending point close to the destination, as shown in **Algorithm 2**. The following notations used in the algorithms are listed below.

**Notation:**

$O$: Origin

$D$: Destination

$E$: Set of all directed edges representing real-world roads

$G$: Directed graph output by Road Network Abstraction algorithm

$N$: Set of nodes (centroids)

$A$: Set of arcs (links)

$\alpha, \beta, \gamma$: Proximity thresholds for searching GPS points close to OD, neighboring nodes, and candidate links, respectively

$M$: One-to-many mapping between node id and node GPS

$T$: A trip trajectory between the origin and the destination

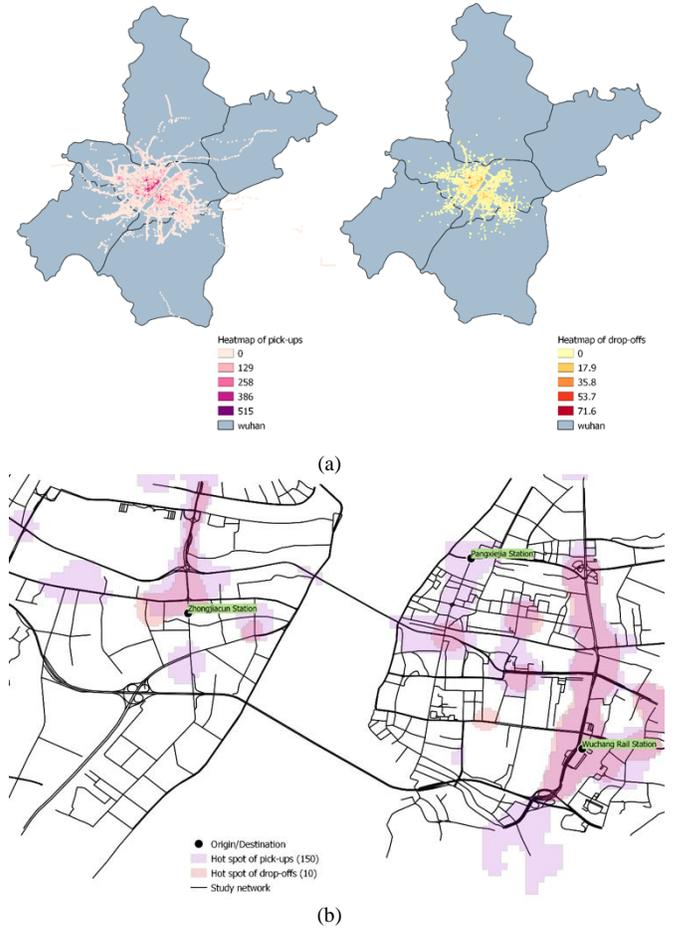

Fig. 6. (a) Heatmaps of taxi pick-ups and drop-offs on May 6th, 2014 in Wuhan, China; (b) Hot spots of taxi pick-ups and drop-offs on May 6th, 2014, and the OD studied in the test

$B$: A "best-fit path" including a sequence of links that best fits a GPS trip trajectory

$h$: Equidistant point (hole) on a road (of a link)

---

### Algorithm 2 Trip Extraction

**Data:** Dataframe *df* contains GPS traces of a taxi, an *origin* and a *destination*
**Result:** All trip trajectories between origin and destination, allowing deviations within the threshold $\alpha$
*curCarId* = -1
**forall** row in *df* **do**
    *carId* = row['Car Id']
    **if** *carId* != *curCarId* **then**
        *inBetween* = False # True if the point belongs to the sub-trajectory between OD
        *pickPasenger* = False # True if the car picks up a passenger
        *havePassenger* = False # True if the car is serving
        *dropPassenger* = False # True if the car drops off a passenger
    **end**
    *p*=(row['lot'], row['lat']) # a point with longitude and latitude information
    **if** distance(*p*, *origin*) ≤ $\alpha$ **then**
        *inBetween* = True
        **if** row["status"] == 1 and not *havePassenger* **then**
            *pickPasenger* = True
        **end**
    **end**
    **if** distance(*p*, *destination*) ≤ $\alpha$ **then**
        *inBetween* = False
        **if** row["status"] == 1 and *havePassenger* **then**
            *dropPassenger* = True

```
            end
        end
        if dropPassenger then
            finish reconstructing route
            record the route information
        end
        if inBetween and pickPassenger then
            havePassenger = True
            start reconstructing route
            pickPassenger = False
            continue
        end
        if havePassenger then
            continue reconstructing route
        end
    end
end
```

### F. Data preparation: road network abstraction

The OpenStreetMap shapefile consists of several features and each represents a real-world road fraction with a list of GPS points. We obtain a directed graph by drawing a directed edge from the first GPS point to the last one for each feature. If a feature has a "oneway" attribute $B$, the corresponding road is bidirectional. Hence, we add an additional directed edge in the opposite direction for such features. Denote the set of all directed edges as $E$. Each edge has two endpoints and a weight equal to free-flow time calculated as $\frac{length}{maxspeed}$, where the $length$ is measured from the feature and the $maxspeed$ is an attribute of the feature denoting the speed limit of the road. The missing values of "$maxspeed$" are filled according to the attribute "$fclass$" (tags for identifying the kind of road).

Considering there are many roads with a short length, especially at the crossroads and roundabouts, we remove such roads (edges) by merging the endpoints close to each other, replacing them with a centroid (see **Algorithm 3**). We define "close" by a threshold $\beta$, say 50 meters. The output of **Algorithm 3** is a directed graph $G = (N, A)$, where $N$ is the set of the centroids (nodes) and $A$ is the set of arcs.

The main idea of the grid method (see geometric hashing [35]) used to speed up in **Algorithm 3** is to first find the boundary of the road network and divide the network into grids of the same size according to the side length of the boundary, then search for the nearest node in the adjacent grids (with Manhattan Distance not exceeding length along one grid). In the specific implementation, each node has a copy in its 8 adjacent grids. Now the nearest node can be searched in one grid, which greatly reduces the search time (brute-force approach must traverse all nodes to calculate distances between them).

---

**Algorithm 3 Road Network Abstraction**

**Data:** $E = \{e \mid e = (point_1, point_2)\}$ contains all edges data. Each edge contains two endpoints and each endpoint contains GPS (lat and lot) information.

**Result:** Output a graph $G = (N, A)$ representing $E$. Merge those endpoints whose distances between each other are less than specific threshold $\beta$ (say 30 meters) to a node in $A$.}

Let $M = \{\}$ be a one-to-many mapping between node id and node GPS.
Let $N = \{\}$, $A = \{\}$ be the set of nodes and arcs.
$i = 1;$
**forall** $e = (point1, point2) \in E$ **do**
　　$NearestNode$ returns the id of nearest node in $M$
　　$n = NearestNode(point_1, M); n_1 = n$
　　**forall** $p \in M[n]$ **do**
　　　　**if** $distance(p, point_1) > \beta$ **then**

```
            n_1 = i; i = i + 1;
            N = N ∪ n_1;
            M = M ∪ {(n_1, point_1)};
            break;
        end
    end
    if n_1 == n then
        M = M ∪ {(n_1, point_1)};
    end
    n = NearestNode(point_2, M); n_2 = n
    forall p ∈ M[n] do
        if distance(p, point_2) > β then
            n_2 = i; i = i + 1;
            N = N ∪ {n_2};
            M = M ∪ {(n_2, point_2)};
            break;
        end
    end
    if n_2 == n then
        M = M ∪ {(n_2, point_2)};
    end
    if (n_1, n_2) ∉ A then
        A = A ∪ {(n_1, n_2)};
    end
end
```

### G. Data preparation: best-fit path

The critical step in the simulation is to reconstruct the actual path of a taxi based on its GPS trip trajectory. However, given OD pairs on the directed graph, there are multiple candidate paths starting from the origin and ending with the destination, even if the nodes and links are not allowed to be accessed repeatedly. To select an optimal path among the candidate paths, we give the definition of the "best-fit path" here.

**Definition 3.1** Given a directed graph $G(N, A)$ and a trip trajectory $T = [p_1, p_2, ..., p_m]$ of GPS points, where $A$ consists of links representing real-world roads on the map. Each $p_i$ is an observed GPS point containing longitude and latitude information. Let $t(a)$ be the tail of link $a$ and $h(a)$ be the head, and $c_a$ is the length of link $a$. A "best-fit path" $B = \{a_1, a_2, \cdots, a_k\}$ $(a_i \in A)$ is a minimum length directed path $argmin_B \sum_{a \in B} c_a : \{h(a_i) = t(a_{i+1}) \; \forall i = 1 \dots k - 1; \exists j = 1, \dots, k : d(p_i, a_j) \leq \gamma \; \forall i = 1, \dots, m - 1\}$. Here $d(p, a)$ is the projection distance between point $p$ and link $a$, $\gamma$ is a pre-defined threshold.

Considering 1) the potential GPS errors that bring troubles to the search of "best-fit path" and 2) relatively large time intervals between adjacent GPS records that can lead to many possible paths, we base our path reconstruction algorithms on two assumptions:

　i.　GPS errors do not exceed the threshold $\gamma$.
　ii.　As a rational man, a taxi driver often locally chooses the shortest path (between the moments when adjacent GPSs are collected).

### H. Data preparation: path reconstruction

There are some difficulties with path reconstruction. First, the GPS points have certain inaccuracies. It's hard to tell which road a taxi is on, simply return the nearest link may not always be correct. Consider the following three cases: 1) the taxi is on a bidirectional road, the brute-force approach returns the link with opposite direction to the correct one; 2) there is a highway right above a road, the brute-force approach returns the highway when the taxi is actually on the road or the other way around; 3) the taxi is close to the intersection of two roads, the brute-force approach returns the link intersected with the correct one. Second, even if we can decide the road information of two consecutive GPS points, the two roads may be disjoint because the interval between the two data points may be long enough to pass multiple roads (it may take only 20 seconds to pass a block). We need to do data imputation. For example, when a taxi is crossing the Wuhan Yangtze River Tunnel, there is no GPS record due to the bad signal. However, we can use the last GPS record before entering the tunnel and the first one after exiting the tunnel to inference the missing roads (tunnel). Last, we have to select the best-fit path from several possible paths for a GPS trajectory.

Considering the first difficulty, we find several link (arc) candidates for each point on a trip trajectory (see **Algorithm 3**). As for the second difficulty, each time we obtain link candidate sets $X$ and $Y$ of adjacent GPS, for each head $x$ in $X$ and for each tail $y$ in $Y$, we find the shortest path from $x$ to $y$ using Dijkstra's algorithm (see **Algorithm 4**). *PunchLine* returns a mapping $M'$ between link id and hole GPS. A hole is one of the equidistant points (including endpoints) on the road represented by a link. *NearestHole* returns the link id and hole GPS of the nearest hole in $M'$. *GetallLinks* returns all links if there exists a hole of the link such that the distance between the hole and $p$ is less than the threshold $\gamma$. The brute-force approach to implement *GetallLinks* runs in O($TE$) time, so we use the grid method to speedup.

Since the maximum time interval is just 2 minutes, we consider the shortest path algorithm is reliable to fill in only a few missing roads. Now with several possible paths for a GPS trajectory, we choose the shortest one as the best-fit path (see **Algorithm 5**). *Shortest*($n_j$, $n$, $B$) is a function that gets the shortest path $B_x$ from $n_j$ to $n$ with the knowledge $B$.

Two steps can be time-consuming. The first is to find candidate links. The brute-force approach searches for real-world roads corresponding to all arcs in the directed graph and calculates the projection distance between each GPS point and each road. The second is that we need to traverse not only all candidate link sets, but also all candidate links in each set, and repeatedly call Dijkstra's algorithm for each combination of candidate links. For the former, we consider drilling equidistant holes for each road and then searching for the road where the nearest hole is located. In this way, we no longer need to calculate the projection distance between a point and a curve (road), but the Euclidean distance between a point and a point instead. Here we also use the grid method mentioned above to speed-up. The only difference is that if nothing returns after searching the grid and the 8 adjacent grids, we continue searching the grids in outer layers until we reach a threshold. For the latter, we also apply some speed-up methods: 1) using tree data structures and dynamic programming; 2) merging duplicate segments of candidate paths.

The experiment parameters are summarized in **Table III**. A plot of the observed path travel time/free flow travel time ratio over the 180 samples is shown in **Fig. 7.**

The next section provides the results of monitoring with the MAIO method and comparison of estimated trip travel times using the inferred capacity dual variables $w_i^*$ to the observed travel times.

| | |
|---|---|
| Number of nodes | 855 |
| Number of links | 2833 |
| Observation period | from 5AM to 9AM on May 6th, 2014 |
| Average time between observations | 5 minutes |
| Number of time intervals | 48 |
| Number of samples observed for OD 1 | 132 |
| Number of samples observed for OD 2 | 48 |
| Number of difference path taken for OD 1 | 53 |
| Number of difference path taken for OD 2 | 29 |
| Average observed path travel time/free flow travel time ratio | 2.72 |

## IV. RESULTS

### A. Simulation of online monitoring

We start with a single OD pair (OD 1: *Zhongjiacun Station* to *Wuchang Rail Station*). The MAIO method was implemented in MATLAB R2017a calling the IBM ILOG CPLEX Optimization Studio v12.8 to solve the inverse shortest path problem in Eq. (8) – (12). Each run to update the dual variables takes less than a second to process.

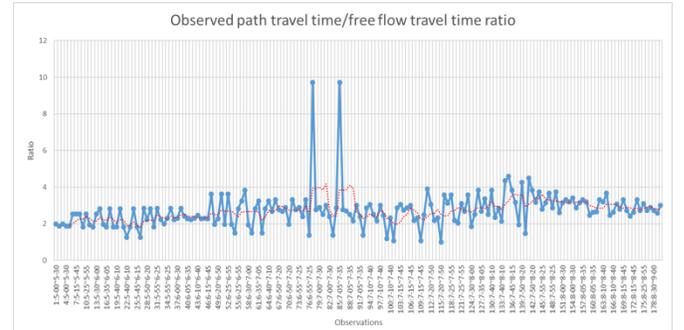

Fig. 7. The observed path travel time/free flow travel time ratio over 180 observed routes from 2 OD pairs.

---

**Algorithm 4 Get Link Candidates**

**Data:** $M$ is a one-to-many mapping between node id and node GPS. $G = (N, A)$ contains nodes and arcs of the whole graph. $T = [p_1, p_2, ..., p_m]$ is a trip trajectory (series of GPS points) of a vehicle. Each $p_i$ contains longitude and latitude information.

**Result:** Output a sequence of candidate link sets $S = [C_1, C_2, ..., C_m]$. Each $C_i$ is a set of candidate link where each link has GPS $p_i$. Link $a$ has $p_i$ if $p_i$ is close to $a$ (within threshold $\gamma$).

Let $S = []$ be the sequence result.

$M' = PunchLine(M, G)$

**for all** $p \in T$ **do**

  $(a, h) = NearestHole(p, M')$;

  **if** $distance(h, p) < \gamma$ **then**

    $C = GetallLinks(p, M', \gamma)$;

  **end**

  **else**

    $C = \{a\}$;

  **end**

  Append $C$ to $S$;

**end**

---

**Algorithm 5 Get Best-fit Path**

**Data:** $G = (N, A)$ contains nodes and arcs of the whole graph, $S = [C_1, C_2, ..., C_m]$ is the sequence of candidate link sets, where $C_i = \{a_{i1}, a_{i2}, ...\}$ is the candidate link set of GPS $p_i$.

**Result:** Output best-fit path $B = (n_1, n_2, ..., n_{k+1})$ passing $T$ where each $n_i \in N$ is a node.

```
for i ∈ [1, m] do
    Let C′ᵢ={ n | n ∈ A[a], a ∈ Cᵢ } be the set of endpoints (nodes) of all links in
    Cᵢ
end
Let S_B = { } be the set of all possible paths
forall n ∈ C′₁ do
    B = (n);
    S_B = S_B ∪ {B};
end
i = 2;

while i ≤ m do
    S′_B = { }    # S′_B is the new possible path
    forall n ∈ C′ᵢ do
        forall B ∈ S_B do
            Bₛ = Shortest(nⱼ, n, B);
            record new B′ = B ∥ Bₛ    # get the shortest path B′ ending with n
        end
        S′_B = S′_B ∪ B′
    end
    S_B = S′_B    # the set of all possible paths that end with points in C′ᵢ
    i = i + 1
end
```

return the shortest path in $S_B$

**Fig. 8(a)** shows the trajectory of the link dual variables (the ones that became binding) as they evolve from one new observation update to the next. The figure illustrates the sensitivity of the method to changes in the network parameters over time, based on the 132 observed individual route choices. There are 409 links traversed by taxis from *Zhongjiacun Station* to *Wuchang Rail Station* in the morning from 5:00 AM to 9:00 AM, and dual variables are positive on 25 links. Those links are highlighted in red in the map, and ones with higher dual variables are labeled.

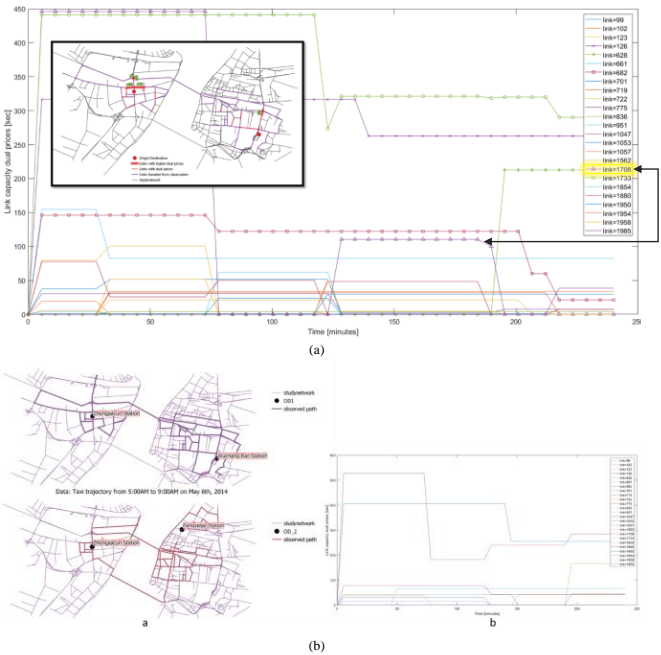

Fig. 8. (a) Trajectories of link dual variables as estimated using Algorithm 2.1 for study network over a 4-hour period; (b) Updated dual variables from new observations as new OD added in the network.

If this result was true, it suggests that link 1708 (road segment near the transport corridor) has the highest dual variable (i.e.

446 sec) before 6:35AM, which means that the link 1708 was the most congested before the congestion was eliminated between 6:35AM and 7:25AM. Then there is a light delay at 110 seconds, and it is not back to 0 again until 7:55AM. This suggests there was an incident in the earlier spike as it was not sustained. On the other hand, link 1733 has a sustained congestion throughout the whole period, suggesting heavy usage under recurrent congestion effect. These results confirm the method from Xu et al. [8] in being able to estimate dual variables (or congestion effects) in real-word urban network that can provide interpretable insights to a decision-maker.

What happens if we sampled from multiple OD paths instead of just one (changing our sampling frame)? We add one more OD pair to observations (OD 2: 48) and re-run the experiment to see how the network state changes. The temporal profile of link dual variables looks quite different. The observed links are mapped on the left in **Fig. 8(b)**, while the link dual variables are updated as shown in the profile on the right of **Fig. 8(b)**. The dual variable for link 1708 drops significantly as compared to the one from monitoring a single OD only. This demonstrates that the effectiveness of the MAIO method depends on effective sampling across different OD pairs to provide more comprehensive coverage over the network. Focusing only on data from a limited set of OD pairs can limit the correctness of the magnitude. As more route observations over different parts of the network are considered, they provide more information about the dual variables, which changes the magnitudes of the other paths that overlap. Ideally, every OD pair should be sampled, but this may not always be possible.

### B. Correlation between observed travel times and online monitoring

Finally, we include a comparison between real travel times and estimated travel times to show the accuracy improvement on estimation. We understand that this is data drawn *only* from two OD pairs, so we do not expect a complete picture. Rather, we want to demonstrate that even with only two OD pair sampling, we can achieve some accuracy in the monitoring for the whole network and this improves upon the accuracy achieved with only one OD pair sampling.

**Fig. 9** shows similarities between estimated travel times (i.e. free flow travel time plus estimated dual variables on traveled links) and real travel times (i.e. the time stamp of last GPS points minus the first in each trip segment) for all observed route choices. There are 180 observations in total, of which 48 observations are from the new OD.

We can draw two conclusions. First, graphically estimations based on the MAIO method using two OD sampling is clearly more accurate than using only one OD sampling. Second, when we compute the correlations between the observed and estimated travel times, we see that the correlation value for the single OD and two OD pairs are 0.23 and 0.56, respectively.

We conduct a hypothesis test on the correlation [46] between the $n = \{132, 180\}$ observations of the estimated travel times and observed travel times for the single OD and two OD scenarios. We test $H_0: \rho = 0$ against the alternative $H_A: \rho \neq 0$ and obtain the following test statistic.

$$t^*_{single\ OD} = \frac{r\sqrt{n-2}}{\sqrt{1-r^2}} = \frac{0.23 * \sqrt{132-2}}{\sqrt{1-(0.23)^2}} = 2.69,$$
$$p = 0.008$$

$$t^*_{two\ ODs} = \frac{r\sqrt{n-2}}{\sqrt{1-r^2}} = \frac{0.56 * \sqrt{180-2}}{\sqrt{1-(0.56)^2}} = 9.02,$$
$$p = 1 \times 10^{-5}$$

The p-values are smaller than the significance level 0.05, we can conclude that the correlations are statistically significant using only 2 OD pairs.

This provides validity that using only samples from only two OD pairs we can get a good picture of the actual network, and it improves significantly (more than doubling the correlation) from one OD pair sampling. This suggests that the MAIO method can provide a good fit to the true observations. This is a statistically cheap method as it does not require forecasting population flows. To elaborate, the results suggest that a practitioner can implement a monitoring system that can observe route choices made along a controlled set of OD pairs over time, and use those results to explain congestion effects throughout the network in real time and evaluate intervention strategies as highlighted in **Fig. 1**.

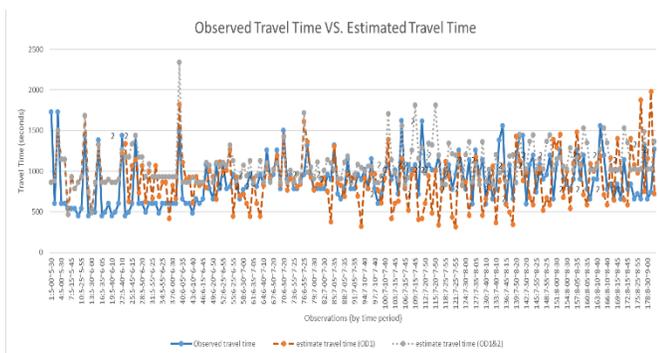

Fig. 9. Estimated travel times and real travel times for 180 observed routes from single and two OD pairs

## V. CONCLUSION

The proposed MAIO model in Xu et al. [8] infers capacity effects throughout a network using only GPS probe samples without the statistically costly step of forecasting population flows. However, the earlier study only provided a theoretical argument and numerical illustration using real data. No validation of the accuracy of the method in monitoring a system is provided. We address this research gap by applying the MAIO method to taxi GPS data in a controlled network setting to simulate an online environment.

Several conclusions are drawn from this empirical validation experiment. Network system attributes like link capacity dual variables can be updated using only samples of individual route observations (e.g., taxi GPS trajectories), without estimating the total link or path flows. This demonstrates that the MAIO method can cheaply monitor a transportation network's system performance over time. The dual variable changes show that the inferences method is indeed sensitive to changes in the system. As traffic increases from 5:00 AM to 9:00 AM in the study period resulting in more spillbacks and incidents impacting link capacities, the set of dual variables steadily increases on average, as shown in **Fig. 8**. The accuracy of the inference is illustrated by the correlation between observed travel time and estimated travel times based on the dual variables updated from the MAIO method. The visual comparison (see **Fig. 9**) indicates the similarities and how they improve after sampling from one OD pair to two. The higher value of correlation for 2 OD pairs shows that the multi-agent IO method performs well in estimating dual variables (or congestion effects) in the form of travel time, and more observations from other OD pairs will only enhance the model performance.

One of the major difficulties in this paper is data processing, that is, how to extract and reconstruct the best-fit paths from raw taxi GPS trajectories. Both GPS errors and the lack of information caused by the large interval between adjacent GPS need to be considered. The path reconstruction process is performed by implementing several algorithms: trip extraction algorithm obtaining trajectories that meet the requirement of our experiments, the road network abstraction algorithm converting the complex map into a directed graph, the candidate edge algorithm finding candidate edges for each GPS points, and the best-fit path selection algorithm applying various pruning techniques and acceleration techniques to efficiently select the best-fit path from a large number of candidate paths for each trajectory.

Future work should implement this system described in **Fig. 1** in a real-world setting using GIS tools and use the monitoring with predefined thresholds to set alerts for dual variables in an online dashboard. Related work can also include monitoring a network before, during, and after a disaster to quantify the impact of dual price increases due to capacity degradation. Since user GPS data may not be freely shared due to privacy concerns, we may experiment with using a blockchain design to anonymize GPS data shared by users or setting up a differential privacy-oriented database [1].

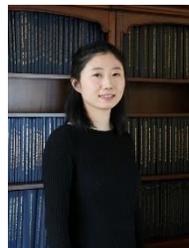

**Susan Jia Xu** received the B.S. degree in civil engineering from Ryerson University, Toronto, Canada, in 2013 and the M.S. degree in transportation system engineering from University of California Irvine, Irvine, CA, USA, in 2015. She got the Ph.D. degree in transportation engineering at New York University, New York, NY, USA.

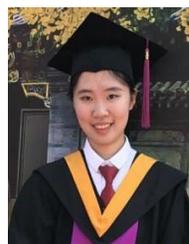

**Qian Xie** is a PhD student at the Department of Civil and Urban Engineering and the C2SMART University Transportation Center in the New York University Tandon School of Engineering. She received her B.Eng. degree in Computer Science from Tsinghua University in 2019.

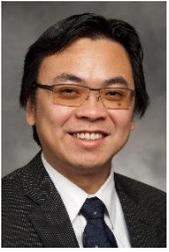

**Joseph Chow** is an Assistant Professor in the Department of Civil & Urban Engineering at New York University Tandon School of Engineering, and Deputy Director of the C2SMART University Transportation Center. His research interests lie in emerging mobility in urban public transportation systems, particularly with Mobility-as-a-Service. He obtained his Ph.D. at UC Irvine in 2010, and a BS and MEng at Cornell University in 2000 and 2001.

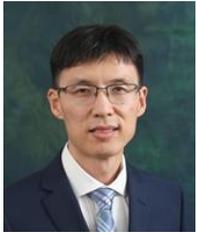

**Xintao Liu** received the B.Eng. degree in survey from Hohai University, China, in 1998, the M.Sc. degree in cartography and GIS from Nanjing Normal University, China, in 2003, and the Ph.D. degree in geoinformatics from the Royal Institute of Technology, Sweden, in 2012.

In 2012, he joined the Department of Civil Engineering, Ryerson University, Canada, where he was a Postdoctoral Fellow in GIS and transportation until 2016. He was a Sessional Lecturer with Ryerson University, since 2015. He is currently an Assistant Professor with the Department of Land Surveying and Geo-Informatics, The Hong Kong Polytechnic University. He is also a PI and a Co-PI of several national projects funded by Sweden, Canada, and Hong Kong. His research interests include GI services and science, urban computing, and GIS in transportation. His research goal is to use the state-of-the-art technologies to advance smart city for a better urban life. He received the Ph.D. Scholarship from Lars Erik Lundbergs. He is a Reviewer of a series of major international journals, such as IJGIS and AAG in his field.